\documentclass[aps,twocolumn,groupedaddress]{revtex4}

\usepackage{epsfig}

\begin{document}
\bibliographystyle{apsrev}


\title{Optical and evaporative cooling of cesium atoms in the
  gravito-optical surface trap}



\author{M.~Hammes}
\author{D.~Rychtarik}
\author{V.~Druzhinina}
\thanks{Present address: 
Physikalisches Institut, Heidelberg University, 69120 Heidelberg, Germany.}
\author{U.~Moslener} 
\thanks{Present address:
Interdisziplin\"ares Institut f\"ur Umwelt\"okonomie, Heidelberg University,
69115 Heidelberg, Germany.}
\author{I.~Manek-H\"onninger}
\thanks{Present address:
Bordeaux University, 33405 Talence, France.}
\author{R.~Grimm}
\affiliation{Max-Planck-Institut f\"ur Kernphysik, 69029 Heidelberg, Germany}


\date{02 May 2000}

\begin{abstract}
\begin{center}
submitted to Journal of Modern Optics,\linebreak 
special issue {\it Fundamentals of 
Quantum Optics V}, edt.\ by F.\ Ehlotzky
\end{center}
We report on cooling of an atomic cesium gas closely above an
evanescent-wave atom mirror. At high densitities, 
optical cooling based on inelastic 
reflections is found to be limited by a density-dependent excess 
temperature and trap loss due to ultracold collisions involving repulsive 
molecular states. Nevertheless, very good starting conditions for
subsequent evaporative cooling are obtained. 
Our first evaporation experiments 
show a temperature reduction from 10\,$\mu$K down to 300\,nK 
along with a gain in phase-space density of almost two orders of 
magnitude.
\end{abstract}

\maketitle

\section{Introduction}
Optical dipole traps based on far-detuned laser light have become 
very popular as versatile tools for the storage of ultracold atomic 
gases and can be employed for a great variety of experiments, 
e.g., on quantum phenomena, precision measurements, 
ultracold collisions and quantum gases \cite{gri00}.
Optical dipole traps have opened up fascinating new experimental 
possibilities not offered by other trapping methods.

A very important advantage compared with magnetic traps is the
fact that a dipole trap can store atoms in any sub-state or 
mixtures of sub-states of the electronic ground state. 
Because of the unusual scattering properties of cesium 
the use of an optical dipole trap may be the only way to achieve 
Bose-Einstein condensation of this interesting atomic species. 
For Cs the quantum-mechanical scattering is resonantly enhanced 
\cite{arn97,hop00},
and binary collisions flipping the spin state lead to severe 
loss from magnetic traps \cite{soe98,arl98,gue98a,gue98b}.
The latter effect, which has been explained  by resonant scattering
in combination with a second-order spin-orbit coupling 
\cite{kok98,leo98}, has so far prevented the attainment of 
a quantum-degenerate gas of cesium atoms.

Inelastic two-body collisions are energetically suppressed 
in the absolute internal ground state of cesium, which is the 
high-field seeking state $F=3,m_F=3$. Atoms in this state 
cannot be trapped magnetically but in an optical trap.
In recent experiments, Vuletic et al.\ have discovered a 
low-field Feshbach resonance \cite{vul99a}, 
which promises an easy experimental way to tune the 
$s$-wave scattering length in a wide range.
With magnetic fields between 17\,G and 30\,G a positive scattering 
length should allow for a stable Bose-Einstein condensate.
 
In our experiments to explore the quantum gas properties of cesium, 
we use the gravito-optical surface trap (GOST) \cite{ovc97}. 
This optical dipole trap allows one to confine a large sample of atoms
in an almost conservative environment with very efficient 
precooling by a Sisyphus-type mechanism. In a second stage,
evaporative cooling is implemented to further increase the 
atomic phase-space density. In this article, 
we summarize the basic properties of the GOST (Sec.~\ref{secGOST}) 
and discuss the limitations of optical cooling at high
densities (Sec.~\ref{secOCOOL}).
We then report our first evaporative cooling results (Sec.~\ref{secEVAP})
and discuss the prospects of future experiments (Sec.~\ref{secOUTLOOK}).

\section{Gravito-optical surface trap}\label{secGOST}

\subsection{Trapping potential and general properties}
\begin{figure}[tb]
\begin{center}\vspace{-7mm}
\epsfig{file=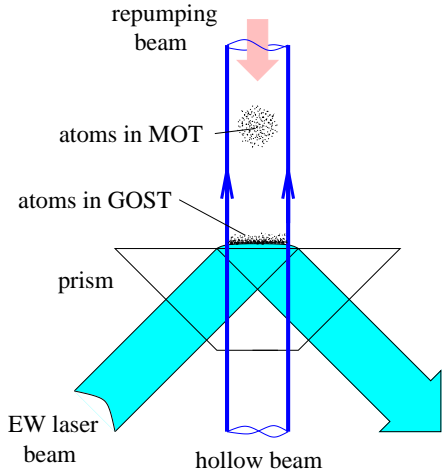, width=6cm}
\end{center}
\caption{Illustration of the gravito-optical surface trap.}
\label{gost}
\vspace{7mm}
\end{figure}

A schematic overview of the geometry of the trap is given in figure
\ref{gost}. 
The GOST is an ``optical mug'', whose bottom 
 consists of an evanescent-wave (EW) atom mirror generated by
total internal reflection of a blue-detuned laser beam from the
surface of a prism, while the walls are formed by an
intense hollow beam (HB) 
which passes vertically through the prism surface. 

The steep
exponential decay  of the EW intensity along the vertical direction and 
the sharp focussing of the hollow beam lead to large intensity
gradients and thus in combination with the blue detuning of both light
fields to a strong repulsive dipole force. We exploit this fact to 
efficiently keep the atoms in the dark inner region of the trap where
the unwanted effect of heating through scattering of trapping light
photons is suppressed. In addition to that, the concept also features
a large trapping 
volume which allows for a transfer of a large number of atoms into
the GOST. Due to the accurate focussing of the hollow laser beam into
a ring-shaped intensity profile \cite{man98}, the
shape of the potential is box-like along the horizontal directions
whereas the combination of gravity and the repulsive wall of the EW
leads to a wedge-shaped potential vertically. 

In thermal equilibrium this geometry allows for a simple description of
the ensemble. The vertical density profile is given by the ``barometric''
equation \begin{equation}n(z)=n_0 \exp(-z/z_0)\end{equation} with
$n_0$ being the peak number 
density and $z_0=k_BT/mg$ the $1/e$-height of the sample.
$m$ denotes the mass of the cesium atom, $g$ the gravitational
acceleration, $z$ 
the vertical coordinate and $T$ the temperature. Assuming a homogeneous 
horizontal density distribution within the hollow beam radius $r_{HB}$ 
this expression can be integrated to express the peak density $n_0$ in 
terms of the atom number $N$ and the temperature $T$,
\begin{equation}\label{density}n_0=\frac{mg}{\pi 
    r_{HB}^2k_B}\,\frac{N}{T}\,.\end{equation}  
The mean density $\bar{n}$ defined as
\begin{equation} \bar{n}\equiv\frac{1}{N}\int d^3r\,n^2({\bf r})\end{equation} 
is exactly half of the peak density ($\bar{n}=n_0/2)$. 
In the GOST potential the scaling of density with temperature
($n\propto T^{-1}$) is somewhat less as compared to a 3D harmonic
oscillator where $n\propto T^{-3/2}$. 

The mean energy per atom is $\bar{E}=5/2\,k_B T$,
of which $3/2\,k_BT$ are kinetic energy and one $k_BT$ potential energy in 
the vertical gravitational field, as can be obtained using the Virial
theorem. In the horizontal degree of freedom the box form of the
confining potential leads to negligible potential energy. 

While the horizontal extension of the
sample is determined by the diameter of the hollow beam (2\,$r_{HB}$),
the height is proportional to the ensemble
temperature with a constant of proportionality of 6.4\,$\mu$m/$\mu$K. At
typical equilibrium temperatures of a few $\mu$K
the $1/e$-height is on the order of a few ten $\mu$m thus leading to a highly
anisotropic ``pancake-shape'' of the ensemble.

\subsection{Experimental realization}
The experimental constituents of the GOST are the evanescent-wave
diode laser, a 
titanium:sapphire laser to create the hollow beam and an additional
diode laser to provide the light for the repumping beam. For EW and
repumping beam we use 
laser diodes (SDL-5712-H1, distributed Bragg reflector), which yield
up to 100\,mW of radiation  
at the cesium D$_2$-line at a wavelength of 852\,nm.

The EW laser is focussed to provide a round spot on the surface of the 
prism with a $1/e^2$-radius of 540\,$\mu$m. The angle of incidence is
$\theta=45.6^\circ$ which is $2^\circ$ above the critical angle of
$43.6^\circ$. This leads to a $1/e^2$ decay length of
$\Lambda=(\lambda/2\pi)(n^2 \sin^2\theta-1)^{-1/2}\approx 500$\,nm, where 
$n=1.45$ is the refractive index of the fused-silica prism.

The EW detuning $\delta_{EW}$ is between a few GHz 
during optical Sisyphus cooling and up to 300\,GHz in the process of
evaporative cooling.
Since atoms are predominantly kept in
the lower hyperfine ground state ($F$=3) in the dipole trap, all
detunings given 
throughout this paper refer
to the transition between this state and the center of the
$^2$P$_{3/2}$ excited-state manifold. 

At $\delta_{EW}/2\pi = 3$\,GHz and a power of 45\,mW the EW creates a 
repulsive
optical potential with a height of $\sim 1$\,mK. The potential barrier 
is reduced 
by typically a factor of two by the attractive
van-der-Waals interaction between the atoms and the dielectric surface 
\cite{lan96a}.

The hollow beam is generated using an axicon optics \cite{man98} to
create a ring-shaped focus of an inner and outer 1/e-radius of
$r_{HB}=260\,\mu$m and $r_{HB}+\Delta r_{HB}=290\,\mu$m,
respectively. It has a power of 350\,mW and its detuning is in the
range between $-0.3$\,nm and $-2$\,nm. The HB provides a potential
barrier on the order of 
$100\,\mu$K height. At this detuning the HB potential is almost
conservative as the photon scattering rates can be estimated to be
on the order of a few photons up to a few ten photons per second.

The repumping beam needed for the optical Sisyphus cooling \cite{soe95}
is resonant  
with the $F=4\rightarrow F'=4$ hyperfine transition of the $D_2$-line
and has an intensity on the order of 1\,$\mu$W/cm$^2$. It is shone on the
trapping region from above (see fig. \ref{gost}).

\subsection{Loading of the trap}
The loading scheme of the GOST goes along standard ways starting from an
effusive atomic beam which is decelerated by a Zeeman slower \cite{inkaDr}.
A magneto-optical trap (MOT) collects about $3\times 10^8$ atoms
under ultrahigh vacuum conditions ($\sim$10$^{-11}$\,mbar). 
The sample is then cooled
to $\sim 10\,\mu$K and spatially compressed 
using a polarization gradient cooling
scheme, in which the MOT laser detuning is increased within 50\,ms
from 3\,$\Gamma$ to 14\,$\Gamma$; here $\Gamma/2\pi=5.3\,$MHz denotes
the natural linewidth. 
After the atomic cloud is shifted from its loading position (3\,mm
above the surface) to a
release position (0.5\,mm) close to the evanescent
wave using magnetic offset fields, the MOT-laser beams are shuttered
and the atoms fall onto the EW-light sheet. As soon as the
near-resonant light fields of the MOT are switched off,
the repumping beam of the GOST is turned on to optically
pump the atoms into the lower hyperfine state ($F$=3), 
and the Sisyphus cooling in the GOST starts. Initially up to
$2\times10^7$ atoms are transfered into the dipole trap and undergo
optical cooling.  

The number of atoms $N$ remaining in the GOST after a variable storage
time is measured by recapturing them into the MOT and taking a 
fluorescence picture using a slow-scan
CCD camera. The integrated fluorescence signal, calibrated with
a more accurate absorption image of the MOT, allows us to determine
$N$ with an estimated uncertainty of less than 50\%.

\section{Optical cooling at high densities}\label{secOCOOL}

Here we discuss the efficient EW Sisyphus 
cooling mechanism and its role as optical precooling stage 
for evaporative cooling in the GOST. We present our observations 
of density-dependent effects that limit EW cooling, namely an excess
temperature and trap loss due to ultracold collisions in the presence
of blue-detuned light.

\subsection{Evanescent-wave Sisyphus cooling}
The optical cooling process is based on inelastic reflections of the
repeatedly bouncing cesium atoms from the evanescent wave 
\cite{soe95,des96}. 
In the
great majority of the bounces the atoms are coherently reflected in
the lower hyperfine state ($F$=3) without any dissipation of kinetic
energy or heating. 
However, occasionally the absorption of an EW photon takes
place and the subsequent spontaneous decay will either have the atom
end up in the $F$=3- or the $F$=4 hyperfine state. A decay into the upper
hyperfine state occurs with a branching ratio of $q$ = 0.25 and due to
the reduced dipole force on this state will lead to a weaker repulsion 
of the atom from the EW and thereby to a damping of the vertical motion.
After leaving the EW the atom is optically pumped back into the lower
hyperfine state. Simple considerations on the balance of cooling
through this process and heating due to photon scattering lead  
to an expression for the equilibrium temperature:
\begin{equation}
T\,=\,\left(\frac{1}{q}\,+\,\frac{1}{q_{\rm
    r}}\right)\,\left(1\,+\,\frac{\delta_{\rm EW}}{\delta_{\rm HFS}}\right)
\,T_{\rm rec}.
\label{coolinglimit}
\end{equation}
The first term in brackets represents the average number of photons
scattered per cooling reflection. $q_r=5/12$ denotes the 
branching ratio of the decay of the $F$'=4 excited state into the
$F$=3 ground state. $\delta_{\rm EW}$ is the detuning of the
evanescent wave and $\delta_{\rm HFS}= 2\pi\times 9.2$\,GHz the
hyperfine splitting of the  
ground state. $T_{\rm rec}= 200\,$nK is the recoil temperature of cesium. The
experimental parameters yield a cooling limit of slightly less than
2\,${\bf\mu}$K. 

\begin{figure}[tb]
\begin{center}
\epsfig{file=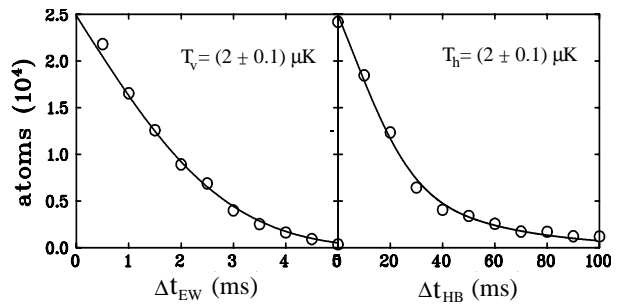, width=8cm}
\end{center}\vspace{-7mm}
\caption{Measurement of the equilibrium temperature for 
$2.5\times10^4$ trapped atoms at a large HB detuning of 2\,nm.
The employed release-and-recapture method allows to 
determine the horizontal and vertical temperatures $T_h$ and $T_v$ 
separately. This is accomplished by 
switching off either the EW or the HB for a short time interval
$\Delta t_{EW}$ or $\Delta t_{EW}$, respectively, 
and measuring the corresponding loss of atoms.
Under equilibrium conditions the sample has an isotropic temperature, 
which in this case ($T=2.0\,\mu$K) 
is very close to the theoretical cooling limit 
of EW Sisyphus cooling.}
\label{tmin}
\vspace{7mm}
\end{figure}

These considerations are
consistent with observed temperatures of 2.0\,$\mu$K in small samples
($N \lesssim 10^5$ atoms) at large hollow beam detunings (see figure
\ref{tmin}).
Dense samples with $N \gg 10^5$ 
atoms show an excess temperature as will be 
discussed in the following section.

The time scale on which the cooling process takes place is on the
order of one second
-- considerably longer than in a MOT. Theoretical
considerations on the 
cooling dynamics \cite{soe95} 
predict an exponential drop of
the vertical temperature in the beginning of the cooling process with a rate
constant  
\begin{equation}
 \beta = 
 \frac{q}{3}\frac{\delta_{HFS}}{\delta_{EW}}
\frac{mg\Lambda}{\hbar(\delta_{EW}+\delta_{HFS})}\,\Gamma\,.
\end{equation}
At $\delta_{EW}=2\pi\times3$\,GHz we find $\beta=1.1$\,s$^{-1}$. 

Cooling of the horizontal motion is facilitated through mixing 
of the vertical and the horizontal degrees of freedom by either
diffuse reflection of atoms from the EW \cite{ovc97,lan96b} or in the
case of dense samples by elastic collisions.
Right after the transfer from the MOT into the GOST, the potential
energy from the fall is gradually converted into thermal energy and
leads to a very hot ($\sim 100\,\mu$K) sample after thermalization
within $\sim0.5\,$s.
The few seconds it takes for the ensemble to reach
the equilibrium temperature are consistent with the calculated cooling 
rate.

To measure temperatures we use a release-and-recapture method
which is accomplished by turning off the EW-potential for a short
duration (few milliseconds) and measuring the remaining fraction of atoms as 
a function of this release time. A fit of a
theoretical model to the data which is based on a Boltzmann
distribution \cite{inkaDr}, yields $T$ as the only fit
parameter. Under the conditions of the experiments reported here the
sample is almost thermalized at any time, so that separate
measurements of the horizontal temperature were not routinely performed.

\subsection{Excess temperature in dense samples}
\label{temp}
\begin{figure}[tb]
\begin{center}
\epsfig{file=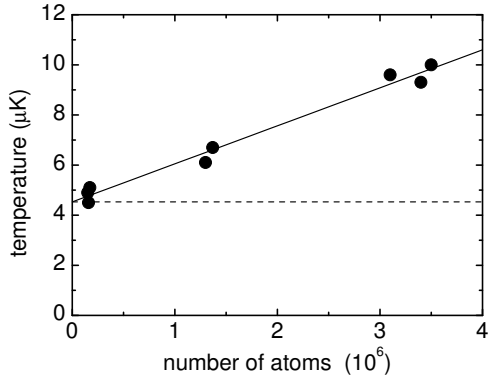, width=8cm}
\end{center}\vspace{-7mm}
\caption{Equilibrium temperature versus atom number for 
$\delta_{EW}/2\pi=6$\,GHz and a HB detuning of 0.3\,nm. The solid line 
represents a linear fit to the experimental data (dots). 
The dashed line indicates the resulting limit temperature 
for very low densities.}
\label{tempverlauf}
\vspace{7mm}
\end{figure}

Measurements of $T$ in large atomic samples yield significantly higher 
temperatures ($\sim$$10\,\mu$K) compared to
what we found in earlier experiments on small samples \cite{ovc97}. 
It turns out that the equilibrium
temperature of EW Sisyphus cooling depends critically on the number of
trapped atoms.  Figure \ref{tempverlauf} shows the strong 
dependence of $T$ on atom number for typical operating conditions of
the GOST. 
The data is reasonably described by a linear dependence
\begin{equation}\label{T(N)}T=T_0+a N\,,\end{equation} 
where $T_0=4.5\,\mu$K is the
limit temperature achieved in small samples and
$a=1.5\,\mu$K$/(10^6$\,atoms) the slope.
The limit temperature $T_0$
can be reduced close to the EW Sisyphus cooling limit (see
eq.~\ref{coolinglimit}) by increasing the detuning of the hollow beam 
(see figure \ref{tmin}), which is easily explained by heating due to
photon scattering. However at large detunings it is not possible to
transfer large numbers of atoms into the GOST so that one has  to find 
a compromise between appreciable atom number and low temperature.

The excess temperature as observed in large samples indicates the
presence of multiple photon scattering. Similar observations are made
in experiments on ``gray molasses'' where one also keeps the atoms
predominantly in the $F=3$ ground state \cite{boi96}.
We find that the slope $a$
decreases with increasing HB detuning. It is however not obvious
that this excess heating solely depends on the hollow beam. The repumping beam 
of EW Sisyphus cooling is likely to contribute to this effect.

In a new experimental setup, in which the HB optics was substantially 
improved to provide aberration-free focussing \cite{ryc00}, we
are now able to trap $10^7$ atoms at a HB detuning of 1\,nm and achieve a 
temperature of $\sim10\,\mu$K. This is better by a roughly a factor 
of two as compared to the conditions of the reported experiments.
The improved setup was used for the evaporation experiments in
Sec.~\ref{secEVAP}.

\subsection{Binary collisions in a blue-detuned light field}
\label{collisions}
\begin{figure}[tb]\label{decay}
\begin{center}
\epsfig{file=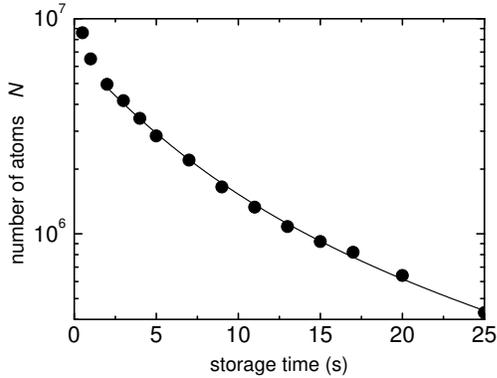, width=8cm}
\end{center}\vspace{-7mm}
\caption{Decay of the number of trapped atoms with increasing storage
time under the same experimental conditions
as in fig.~\protect\ref{tempverlauf}.
The measured data points (dots) are well fitted by a theoretical 
model that takes into account loss due to binary collisions. The fit
(solid line) yields $\alpha = 0.03\,$s$^{-1}$ and 
$\beta=0.9\times10^{-12}\,$cm$^3/$s for the rate coefficients.}
\vspace{7mm}
\end{figure}

Lifetimes of samples of more than $10^6$ atoms were found to be on the
order of 5 
to 10 seconds and clearly indicated the presence of a strong
nonexponential contribution to the decay (see figure \ref{decay}). We
model the decay 
assuming the presence of a two-body loss process in the standard loss
rate equation \cite{wei99}
\begin{equation} 
\dot{N}=-\alpha N -\beta \bar{n}N\,.
\end{equation} 
To solve this equation one has to consider the complete dependence of
the mean density
$\bar{n}$ on the atom number $N$, which is also influenced by the
observed excess temperature. Since cooling and thermalization takes
place on a time scale much shorter than the decay one can assume
stationary conditions for $T(N)$ according to eq. \ref{T(N)}.
Using this and
equation \ref{density} one obtains \begin{equation}\bar{n}=\frac{mg}{2 \pi 
    r_{HB}^2k_B}\,\frac{N}{T_0+a N}\,. 
\end{equation}
Solving the differential equation one can almost perfectly fit the
experimental data and extract the quadratic loss coefficient $\beta$
as a fit parameter.
 Figure \ref{decay}
demonstrates the good agreement between the model and the
measurement.  
We find $\beta$ to be on the order of
$10^{-12}$\,cm$^3$/s, which is about an order of magnitude less as
compared to radiative escape in a MOT \cite{wei99}. 

A possible mechanism for the observed loss is based on the excitation of a
pair of colliding cesium atoms into a repulsive molecular state
\cite{bal94,hof96,bur96,vul99b}. The
resonant dipole-dipole interaction splits the excited molecular state into an
attractive and a repulsive branch. While the attractive branch gives
rise to loss processes like radiative escape and photoassociation in
red-detuned light, the repulsive branch becomes relevant in
blue-detuned light.  
When a cesium pair is excited into a repulsive molecular state at the 
classical
Condon point it is
accelerated along the potential curve and obtains a kinetic energy
equivalent to the detuning of the exciting light field with respect to 
the atomic transition. For a dipole trap the resulting energy gain is
in general much larger than the trap depth and leads to an ejection of both
colliding atoms from the trap. 
According to a simple
semi-classical model the corresponding rate coefficient $\beta$ should scale
with the laser detuning and intensity as $I/\delta^2$ as long as the
intensity is low enough to avoid optical shielding effects
\cite{bal94}.

In a first set of measurements to investigate this loss process we obtained 
decay curves at different detunings of the EW-laser field and
extracted the rate coefficient $\beta$. In a gravito-optical trap the
mean EW dipole potential experienced by the bouncing atoms is given by
$\bar{U}_{dip}=mg\Lambda/2$, independent of intensity and
detuning. Therefore the mean intensity $\bar{I}$ to which an atom is exposed 
is proportional 
to the detuning $\delta_{EW}$. Thus the expected scaling
of the rate coefficient is $\beta\propto \bar{I}/\delta_{EW}^2\propto
1/\delta_{EW}$.  
Measurements in the detuning range between 1\,GHz and 7\,GHz \cite{ham99}
indeed showed the expected $1/\delta_{EW}$-behaviour.
This confirms that collisions in the evanescent wave can explain the
observed loss. 

As a practical consequence of this fact we can minimize trap loss during
optical cooling by
switching the EW detuning to higher values shortly after loading. Only 
in the initial cooling phase low detuning is required
to provide a sufficiently large potential barrier. Therefore, in
experiments aiming at high densities, the detuning is routinely
switched from 3\,GHz to about 7\,GHz after 0.5\,s.

\begin{figure}[tb]
\begin{center}
\epsfig{file=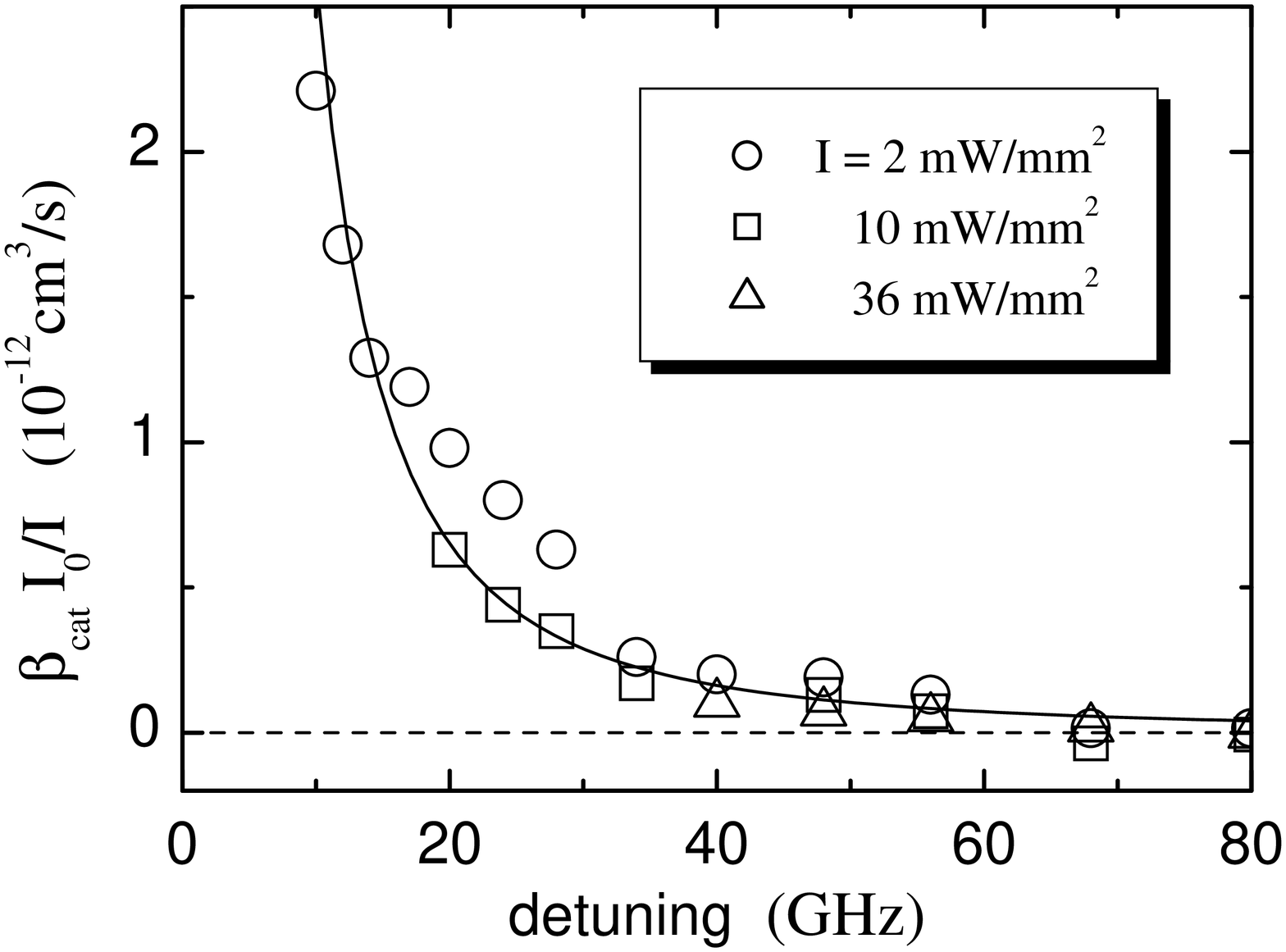, width=8cm}
\end{center}\vspace{-7mm}
\caption{Rate coefficient $\beta_{\rm cat}$ describing the loss induced
by an additional, blue-detuned ``catalysis'' laser. The experimental
data are taken for three different intensities 
and, in order to check the expected proportionality to
$I/\delta^2$, are scaled to the lowest intensity 
$I_0 = 2\,$mW/cm$^2$.
The resulting $1/\delta^2$-behavior is indicated by the solid fit
curve. The deviations of some data around 20\,GHz is likely due to
drifts of the experimental conditions.}
\label{katlaser}
\vspace{7mm}
\end{figure}

To prove that excitation into repulsive molecular states
leads to trap loss we use an additional blue-detuned ``catalysis'' 
laser to induce collisional trap loss. In this way this process can be
studied independently of the parameters of the trapping potential.
 The rate coefficient $\beta_{\rm cat}$ was
measured for different values of detuning  and intensity of the
catalysis laser. In additional temperature measurements we ensured
that heating due to photon scattering from the catalysis laser was
negligible for all combinations of intensity and detuning.
Data was collected at different catalysis
laser intensities. The results for $\beta_{\rm cat}$ shown in 
figure \ref{katlaser} agree with the expected
scaling behaviour $I/\delta^2$. 

As an important conclusion of these results, trap loss by binary
collisions involving repulsive excited molecular states plays a 
significant role in optical cooling. Nevertheless, a dense sample
of many atoms at high densities can prepared. If the evanescent
wave is then operated at very large detunings, a conservative 
trapping potential is obtained with very low photon scattering and
strongly suppressed trap loss, i.e.\ a very good starting point for
evaporative cooling.

\section{evaporative cooling}\label{secEVAP}
The GOST offers favorable conditions to implement forced evaporative 
cooling \cite{ket96} with the prospect to attain quantum-degeneracy of 
cesium or a two-dimensional quantum gas. In contrast to red-detuned 
dipole traps used for evaporation experiments \cite{ada95,eng00}, 
the spatial compression of the cold sample essentially results from
gravity and is thus not affected when the optical potentials 
are ramped down. Moreover, many more atoms are initially loaded into
the GOST as compared to typical red-detuned traps.
Here we describe our first experiments demonstrating the feasibility 
of efficient evaporation in the GOST.


In our evaporation experiments we operate the trap at a HB 
detuning of $-1$\,nm and keep the EW parameters as in the measurements 
described above. In two seconds of optical cooling we prepare
$N=10^7$\,atoms at a temperature of $T=10\,\mu$K, and a peak density of
$n_0=6\times10^{11}\,$cm$^{-3}$.  
For the unpolarized sample in the seven-fold degenerate $F=3$ ground
state this corresponds 
to a peak phase-space density of $D=n_0 \lambda_{dB}^3/7\approx10^{-5}$
where $\lambda_{dB}=h/\sqrt{2\pi mk_BT}$ is the thermal de-Broglie
wavelength. Elastic collisions take place at a rate on the order of
$50$\,s$^{-1}$ and, considering the resonant scattering of cesium
\cite{arn97}, lead to a thermalization time of about 200\,ms.

To implement forced evaporation we lower the EW potential by ramping
up the EW detuning. This simultaneously reduces heating due to photon
scattering 
and suppresses loss through the collision mechanism 
described above.  
Within $4.5$ seconds the EW detuning is increased exponentially from
initially 
7\,GHz up to 250\,GHz. 
This is accomplished by rapid mode-hop free temperature
tuning of the EW diode laser. In the last $2.5$ seconds of the evaporation
ramp the intensity of the hollow beam is reduced from $350\,$mW to
$11\,$mW in order to reduce possible heating by residual light in the dark
center of the hollow beam. The contribution of the HB potential ramp 
to the evaporation remains very small.

\begin{figure}[tb]
\begin{center}
\epsfig{file=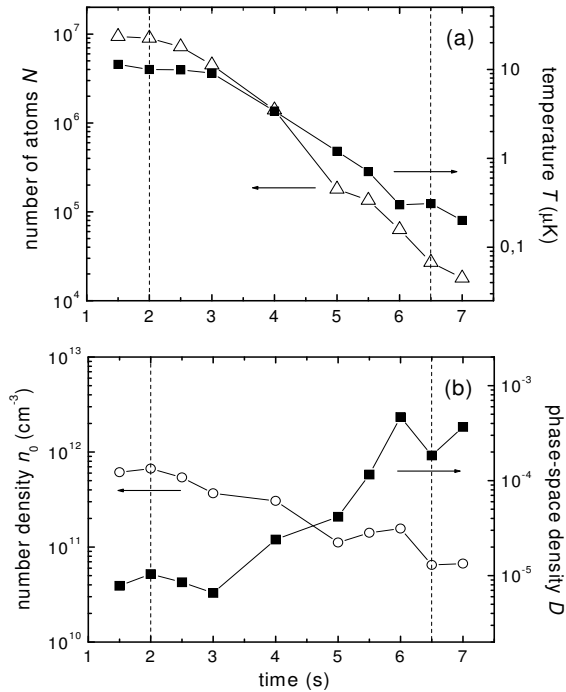, width=8cm}
\end{center}\vspace{-7mm}
\caption{Evaporative cooling in the GOST. The detuning ramp of the EW 
starts after two seconds of optical cooling and ends 4.5\,s later, 
as indicated by the vertical dashed lines. 
The evolution of the number $N$ and the temperature $T$ 
of the trapped atoms is shown in (a), the corresponding behavior 
of peak density $n_0 \propto N/T$ and phase-space density 
$D \propto N/T^{5/2}$ is displayed in (b).}
\label{evapAB}
\vspace{7mm}
\end{figure}

The experimental results are shown in figure \ref{evapAB}. 
About 1\,s after starting the exponential ramp, 
the temperature begins to drop [filled squares in (a)]. 
At the end of the ramp, it has reached $T \approx 300$\,nK. 
This decrease of $T$ by about 1.5 orders of
magnitude is accompanied by a decrease of the particle number $N$ 
[open triangles in (a)] from
$10^7$ down to $\sim$3$\times10^4$, i.e.\ about 
2.5 orders of magnitude.

Although the number density $n_0 \propto N/T$ [open circles in (b)] 
decreases by about one order of magnitude,
the phase-space density $D \propto n_0 T^{-3/2} \propto N T^{-5/2}$
[filled squares in (b)] shows a substantial increase by 1.5 orders of 
magnitude. At the end of the ramp, we obtain a phase-space density of
$\sim$$3 \times 10^{-4}$. 

In the regime of resonant elastic scattering ($T \gtrsim 1\,\mu$K 
\cite{hop00}), the relevant cross section scales $\propto T^{-1}$.
In the GOST potential, this leads to a scaling of the elastic 
scattering rate and the thermal relaxation rate $\propto N T^{-3/2}$.
Therefore the elastic scattering rate is almost constant for the
conditions of our experiments. However, no runaway regime is 
reached.

An obvious problem in our evaporation scheme is that, for the applied 
exponential 
ramp, it takes about one second until the EW potential barrier 
becomes low enough to start the evaporation. 
Up to this point already about 50\% of the particles are lost, 
presumably by
the collisional mechanism investigated in Sec.~\ref{collisions}.
After the corresponding initial loss of phase-space density the
later evaporation then leads to a gain of almost two orders of 
magnitude. This clearly shows that the potential of evaporative 
cooling in the GOST is much larger than we could demonstrate 
in our first experiments discussed here. 

Although the reported evaporation results are still quite preliminary 
they already show that efficient evaporation is
attainable in the GOST by ramping down the optical trapping 
potential. Substantial improvements can be expected by an optimized 
evaporation ramp of the EW in combination with a simultaneous 
evaporation through the hollow-beam potential.

\section{Conclusions and outlook}\label{secOUTLOOK}

We have investigated optical and evaporative cooling in the 
gravito-optical surface trap. 
At high densities, optical cooling by inelastic
reflections from the evanescent-wave bottom of the trap was
found to be limited by an excess temperature, which we interpret 
as a result of multiple photon scattering. In addition, a loss
process is induced by the blue-detuned trap light. Colliding atoms
are excited into a repulsive molecular state which is followed 
by an energy release into the relative motion. Nevertheless, very
good starting conditions are obtained for evaporative cooling.
By reducing the trapping potentials we have cooled the sample
down to a temperature of 300\,nK and obtained a phase-space
of $3\times10^{-4}$.

So far all our evaporation experiments have been performed with
unpolarized atoms in the seven-fold degenerate $F=3$ ground state.
Dramatic improvements can be expected by polarizing the atoms into
the absolute ground state $F=3, m_F=3$. 
The phase-space density can be increased by a factor
of seven, and by producing identical bosons 
the elastic collsion rate goes up by nearly a factor of two.
Moreover, Feshbach tuning can be applied to modify the
$s$-wave scattering length \cite{vul99a}. We have already performed 
first experiments, in which we have demonstrated that atoms the GOST
can be optically pumped into the state of interest \cite{polarizing}.

The spatial compression of the atomic gas can be 
enhanced by adding a far red-detuned laser beam to the GOST.
If such a beam propagates vertically in the center of the hollow beam
it provides an additional potential well for transverse confinement.
As an example, a 500-mW beam from a compact Nd:YAG laser 
focused to a 1/e-diameter of 0.1\,mm already provides 
a dipole potential $U_{\rm red}$ of $\sim$3$\mu$K depth. 
In the GOST such an additional well 
can be loaded by elastic collisions, which in thermal equilibrium 
would result in a peak density enhancement of 
$\exp(|U_{\rm red}|/k_B T)$ along with a corresponding
increase in phase-space density \cite{sta98}.

Another interesting option with an additional far red-detuned laser 
beam is to create a double-EW trap as suggested in Ref.~\cite{ovc91}.
The combination of a repulsive blue-detuned EW in combination with
an attractive red-detuned EW of much larger decay length would 
allow to create a wavelength-size potential well very close to the
dielectric surface. In such a scheme, a situation can be realized
where only one vertical bound state exist. The situation would then
be similar to atomic hydrogen on liquid helium, for which a 
two-dimensional quantum gas has been reported \cite{saf98}. 
The realization of
such a system with alkali atoms could provide much more insight 
into the physical behavior of such a 2D gas.

In addition to such experiments on quantum gas properties, 
the GOST also represents a very promising source of ultracold atoms 
for experiments related to, 
e.g., atom interferometry \cite{szr96}, 
atom-surface interactions \cite{cot98,mar00},
and quantum chaos \cite{sai98}.

\section*{Acknowledgments}
This work was supported by the Deutsche Forschungsgemeinschaft in the
frame of the Gerhard-Hess-Programm. We thank D.\ Schwalm for continuous
support and encouragement. One of us (V.D.) acknowledges a fellowship
by the Konrad-Adenauer-Stiftung.

\pagebreak


\begin{thebibliography}{XX}
\bibitem{polarizing}
For optical Zeeman pumping we use
a weak circulary polarized beam
resonant with the $F=3\rightarrow F'=3$ transition and 
aligned parallel to a static magnetic field of a few Gauss.
In addition the usual GOST repumping light with mixed polarization 
provides hyperfine repumping via the $F=4\rightarrow F'=4$ transition. 
In this scheme the state of interest ($F=3, m_F=3$) is the 
only dark state and thus efficiently populated.
In a Hanle-type experiment \cite{ryc00} 
we have observed that the pumping beams lead to very weak heating 
in the presence of the longitudinal magnetic field. With a transverse
magnetic field destroying the dark state, strong heating was observed.
This confirms the predominant population of the dark state.
\end{thebibliography}
\end{document}